\documentclass{revtex4}
 
\usepackage{graphicx}
\usepackage{dcolumn}
\usepackage{bm}

\begin{document}

\title{Intermittency and scale-free networks: \\
a dynamical model for
human language complexity} 

\author{Paolo Allegrini$^{1}$, Paolo Grigolini$^{2,3,4}$, Luigi
Palatella$^{2}$}

\affiliation{ $^{1}$Istituto di Linguistica Computazionale del
Consiglio Nazionale delle Ricerche, Area della Ricerca di Pisa, via
Moruzzi 1, San Cataldo, 56010, Ghezzano-Pisa, Italy}

\affiliation{ $^{2}$Dipartimento di Fisica dell'Universit\`a di Pisa, via
Buonarroti 2, 56127 Pisa Italy.}

\affiliation{ $^{3}$Center for Nonlinear Science, University of North
Texas, P.O. Box 311427, Denton, Texas 76203-1427 }

\affiliation{ $^{4}$Istituto dei Processi Chimico Fisici del CNR
Area della Ricerca di Pisa, Via G. Moruzzi 1,
56124 Pisa, Italy}

\begin{abstract}
In this paper we try to model certain features of human language
complexity by means of advanced concepts borrowed from statistical
mechanics. We use a time series approach, the diffusion entropy method
(DE), to compute the complexity of an italian corpus of newspapers and
magazines. We find that the anomalous scaling index is compatible with
a simple dynamical model, a random walk on a complex scale-free
network, which is linguistically related to Saussurre's {\em
paradigms}.  The network complexity is independently measured on the
same corpus, looking at the co-occurrence of nouns and verbs. This
connection of cognitive complexity with long-range time correlations
also provides an explanation for the famous Zipf's law in terms of the
generalized central limit theorem.
\end{abstract}

\maketitle
\section{introduction}

This introduction is divided into two parts, the former being devoted
to a general discussion concerning the challenging issue of the search
for crucial events, in general. The second is of linguistic interest. This
division reflects the twofold purpose of this paper. In fact, the main 
result of this paper is the foundation of the Zipf' law, a
subject of theoretical interest to understand the origin and evolution
of language \cite{sole}. The foundation of this important
empirical rule is here discussed from a special perspective. This is
the perspective of complexity conceived as a new condition of
physics. Of course, this condition is not expressed by regular
dynamics, and it is not expressed by thermodynamics, either. This is
rather a regime of transition from dynamics to thermodynamics lasting
for a virtually infinite time, so as to realize a new legitimate state
of physics, affording a perspective valid also in the case of
non-physical systems. A paradigm for this condition is given by the
renewal theory processes \cite{cox}. 
Thus, we make use of the renewal theory 
to define the concept of crucial events.

In conclusion, the first purpose of this paper is to contribute the progress
of a statistical method for the search of crucial events, using written texts
as an illustration of the method. On this important issue we do not reach
conclusive results but the formulation of a conjecture for future
work. 
However, we note that the linguistic results rest on a general 
approach to the detection of rare and significant events. This 
general approach to the detection of crucial events might yield 
useful applications to many fields, from the early diagnosis of 
diseases to the war against terrorism. Thus, we devote the first part 
of the introduction to the definition of crucial events. 
The second part of the introduction serves the
purpose of introducing the reader to the ideas and jargon of linguistics, so
as to fully appreciate the main result of this paper, which is, in fact, of
linguistic interest.

\subsection{Definition of crucial  events}
Let us consider a generic time series. This is a sequel of values occurring at
different times and mirroring the properties of a given complex system, this
being either the heart-beating \cite{memorybeyondmemory}, seismic
movement \cite{earthquakes}, or a written text, as in this paper. Let us label,
or mark, some of these values with a given criterion, which depends, of
course, on the conjectures that we make about the process under study.
Labeling a value implies that we judge it to be significant. Thus, it is
evident that the more we know about the process, the more plausible the
conjecture behind the labeling criterion adopted. To give some illustrative
example, let us mention some of the criteria recently adopted. In the case of
heart-beating \cite{memorybeyondmemory}, we are observing the time distances
$T$ between a R peak and the next, which for a very large numbers of beats $t$
becomes equivalent to a continuous function, $T(t)$, of a continuous variable
$t$. We divide the ordinate axis into small interval of size $s$, and we
record, or mark, the beats at which $T(t)$ moves from one strip to another. In
the case of earthquakes, we label only the seismic fluctuations of intensity
larger than a given threshold \cite{earthquakes}. Finally, in the case of a
written text, as explained later in this paper, we mark the salient words,
these being the words that in a given text appear with a frequency larger than
in a suitable reference text.

Now we have to address the important issue of establishing whether the labeled
values are crucial events, or not. This, in turn, requires that we define what
we mean by \emph{crucial} event. Let us define first the crucial events by
mean of their statistical properties. Then we shall explain why events with
these statistical properties must be considered crucial. We use the
prescriptions illustrated in the fundamental paper of Ref. \cite{giacomo}. We
assign the symbol $1$ to all the marked values and the symbol $0$ to the
others. Then we convert the time series into a set of random walks with the
method detailed in Ref. \cite{giacomo}. We use the walking rule that the
walker makes a step ahead by the quantity $1$ any time we meet a marked value,
namely, the symbol $1$. When we meet the symbol $0$ the walker remains in a
state of rest. The marked values are much less numerous that the unmarked
values. Thus, we shall have many $0$'s and a few $1$'s. A sequence of many
$0$'s between two $1$'s is termed a laminar region. It is important to evaluate
the time distance between one marked value and the next, which is equivalent
to determining the distribution of the time lengths of the laminar
regions. This will result in a time distribution $\psi(t)$. Let us assume that
the waiting time distribution $\psi(t)$ is an inverse power law with index
$\mu$. The Poisson case corresponds to $\mu = \infty$.  Then we study the
resulting diffusion process and we establish its scaling parameter
$\delta$. If the index $\mu$ fits the condition $2< \mu < 3$ and it is
related to the scaling coefficient by the relation $\delta = 1/(\mu-1)$, or,
if the scaling index fits the condition $\mu > 3$ and the scaling coefficient
has the ordinary value $\delta = 0.5$, the labelled values are crucial events.

Let us explain the physical motivation for this definition.
First, meeting a $0$ is not significant because the $0$'s
are closely correlated and meeting one of them implies that that we shall meet
many other $0$'s, before finding a $1$ again. In a sense the $0$'s are driven
by the $1$'s. A given $1$ is the beginning of a laminar region and,
consequently, of a cascade of $0$'s.  On the other hand, if $\delta$ and $\mu$
violate the relations for the $1$'s to be crucial, there might exist a
correlation between two different laminar regions.  This suggests that the
$1$'s might be part of a predictable cascade of some other genuinely 
crucial events,
not yet revealed by the statistical analysis, and that our conjecture is not
correct.

We have to explain why $\mu < \infty$ is an important condition for our
definition of crucial events. We limit ourselves to noticing\cite{gerardo}
that $\mu = \infty$ is equivalent to setting Poisson statistics,
implying, in turn, a memoryless condition that seems to be incompatible with
the labelled values to be crucial.  This is a poorly understood property, in
spite of the fact that 32 years ago Bedeaux, Lindenberg and Shuler
\cite{katja} wrote a clarifying paper on this subject.  We refer the reader to
this fundamental paper\cite{katja} to understand the reasons why we imagine
the crucial events to be incompatible with Poisson statistics.  Actually, the
labelled values seem to become crucial when the diffusion process they
generate is anomalous, namely $\delta > 0.5$, which implies $\mu < 3$. It has
to be pointed out that $\mu< 2$ implies a condition where the mean length of
the laminar region is infinite. This means that in this region the process
becomes non-stationary, due to the lack of an invariant measure
\cite{massi}. From an intuitive point of view, the origin of non-stationary
behavior is as follows. If we keep drawing random number from the distribution
$\psi(t)$, the mean value of $t$ tends to increase with the number of
drawings. This is so because, if it did not increase the overall mean value
would be finite, in conflict with the fact that $\mu < 2$ produces an infinite
mean value. Thus the condition $\mu = 2$ is the border between the stationary,
$\mu > 2$, and the non-stationary condition, $\mu < 2$. As we shall see, this
border has an interesting meaning for linguistics.

It would be an interesting issue to assess whether crucial events can be
located in this region. We limit ourselves to noticing that the results of the
statistical analysis of time series seem to denote that complex systems are
characterized by values of $\mu$ very close to the border without ever
entering the non-stationary dominion $\mu < 2$.  In conclusion, with the
prescription of Ref. \cite{giacomo} we are in a position to assess if the
marked values correspond to crucial events, or do not.

\subsection{The meaning of Zipf's law in linguistics} 

\begin{figure}[!h]
\begin{center}
\includegraphics[width=10 cm]{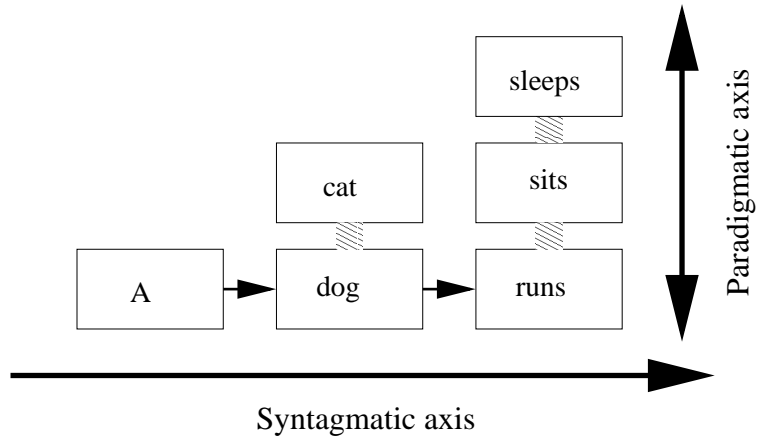}
\caption{\label{saussurre} Saussurre's dimensions. In this
example the first position in the syntagmatic axis is an article,
the second a noun and the third a verb in the third person.
Notice the semantic restriction on the paradigmatic possibilities.}
\end{center}
\end{figure}

Semiotics studies linguistic signs,
their meanings, and identifies the relations
between sign and meaning, and among signs.
The relations among signs (letters, words), are
divided into two large groups, namely the
syntagmatic and the paradigmatic, corresponding to 
what are called Saussurre's dimensions \cite{silverman}. 
A clearcut definition of the two dimensions is outside the scope of
this paper, but one can grasp an understanding of them by looking at
Fig. \ref{saussurre}. The abscissa axis represents the syntagmatic
dimension, while the ordinate axis represents the paradigmatic
one. Along the abscissa certain grammatical rules pose constraints on how
words follow each other. We can say that this dimension is  temporal, with a casual order. An article (as ``a'' or ``the''), 
for instance, may be followed by
an adjective or a noun, but not by a verb of finite form. At a larger
``time-scale'', {\em pragmatic} constraints rule the succession of concepts,
to give a {\em logic} to the {\em discourse}. 
The other axis, on the other hand,
refers to a ``mental'' space. The speaker has in mind a repertoir of
words, divided in many categories, which can be hierarchically
complex and refer to syntactical or semantic ``interchangeability''. 
Different {\em space}-scales of word paradigms can be associated to 
different levels of this
hierarchy. After an article, to follow the preceding example,
one can choose, at a syntactical level, among all nouns of a
dictionary. 
However, at a deeper level, semantic
constraints reduce the avalaible words to be chosen. For instance,
after ``a dog'' one can choose any verb, but in practice 
only among verbs selected by semantic constraints
(a dog runs or sits, but does not read or smoke). The sentence 
``a dog graduates'', for instance, fits 
paradigmatic and syntagmatic rules behind Fig. 1,
but the semantics would in general forbid the production
of such a ``nonsensical'' sentence.

The two dimensions are therefore not quite orthogonal, and connect 
at a cognitive level.
The main focus of this paper is to show that this connection,
which is well known at a short scales, is in fact reproduced
at all scales. We will also show that both dimensions are
{\em scale free} and that the very complexity of linguistic structures in
both dimensions can be taken into account in a unified model, 
which is able to explain most statistical features of
human language, including, at the largest scales, 
the celebrated Zipf's law \cite{zipf}.

Zipf's law, which relates the rank $r$ of a words to its frequency $f$
in a corpus, obviuosly points to constraints on the frequencies of
words. Remarkably, this does not mean that the probability of a word
is actually defined. In fact, a word may have a small or large
frequency depending on the genre of the {\em corpus} (i.e. a large
collection of written text) under study, and even two extremely large
corpora of the same type fail in reproducing the same word
frequencies. It is however remarkable that the occurrence of words is
such that {\em for any corpus and for any natural language} the same universal property
emerges, this being that there are only a few frequent words and there are, on the contraty, many words that are encountered, for instance, only  once or twice. To be more precise, let us define
word rank $r$, which is a property depending on the corpus adopted, as
follows. One assigns rank 1 to the most frequent word, rank 2 to the
second frequent one, and so on. Each word is uniquelly associated to a
rank, and, although this number varies form corpus to corpus, one
always finds that

\begin{equation}\label{zlaw}
f \propto \frac{1}{r},
\end{equation}
We shall derive this law from a complexity model of language, along the
following lines.  We recognize that there is no experimental 
linguistic evidence that {\em word frequencies
tend to well-defined probabilities}. However, we adopt a probabilistic approach, namely,
we assume that there exists, for a randomly selected word, the probability of having the
frequency $f$, denoted as $P(f)$. Operatively, we
measure $P(f)$ by counting how many words have a certain frequency $f$. In
Section III we show that a $P(f)$ compatible with the Zipf's law
can be directly derived from the model proposed herein, thus providing
 a solid experimental support to our model. In order to fulfill our
program, we shall assume a long-range statistical mutual independence for the
occurrence of different {\em concepts}. This hypothesis is
appealing, since this assumption means that every and each
occurrence of a concept makes entropy increase, and therefore we
identify the mathematical {\em information} (i.e. entropy) with the
common-sense information (i.e. the occurrence of concepts). Unfortunately,
a concept is not, {\em a priori}, a well defined quantity. 
Herein we shall assume
that concepts are represented by words (or better by lemmata) or by
groups of semantically similar words or lemmata\footnote{A lemma
is defined as a representive word of a class of words, having
different morphological features. For instance the word ``dogs''has
lemma ``dog'', and word ``sleeping'' has lemma ``sleep''.}.

Because of the mutual independence among different concepts, we can extract
from a single corpus as many ``experiments'' as the number of concepts. For
each experiment we select only one concept and we mark the occurrence of the
selected word or group of words corresponding to this concept.  We shall use
an advanced time-series analysis to study the statistical properties of the
dynamical occurrence of the markers. In fact, as mentioned in the first part
of this introduction, we adopt the DE method \cite{giacomo}, which has been
proved by earlier investigation \cite{memorybeyondmemory,earthquakes} to be an
efficient way to detect the statistics of crucial events. We shall discover
the important result that the adoption of DE method allows us to assess
whether a given set of ``markers" can be identified with the crucial events or
the crucial events are still unknown and we must consequently search for a
different set of markers.

The results of the experiments will lead to a language model reproducing the
language ``complexity'', namely a combination of order and randomness at all
scales.  The overall dynamics, given by the flow of concepts over time,
resemble the dynamics of intermittent dynamical systems, like the Manneville's
Map \cite{manneville}: These systems have long periods of quiescence followed
by bursts of activity. This variability of waiting times between markers of
activity is responsible for long-range correlations.

A further aspect of the paper is the connection between space and time
complexity, and its application to linguistics. We shall assume that atomic
concepts exist and represent nodes of a complex network, connected by arcs
representing, when existing, semantic {\em associations} between a concept and
another. We assume that our markers are actually defined as a group of
neighbor nodes. We then assume that language is the result of a random walk,
namely a random ``associative'' travel from concept to concept. The scale-free
properties of this network, independently measured by several research groups,
including ours, provides a bridge to understand the intermittent dynamics
earlier described. In this unified model the network is a representation of
Saussurre's paradigms, whose complexity mirrors the syntagmatic one in the
asymptotic limit.

\section{DE and concepts} 

Several papers have recently proposed a new way of detecting long-range effects of non-Poisson statistics  \cite{giacomo,giacomo2,earthquakes, nicola,giulia1}. In short, as mentioned in the first part of Section 1, we define a set of ``markers'' among the symbols or values of a time series, and we evaluate numerically the probability of having a number $x$ of markers in a window of length $t$, denoted as $p(x;t)$. This statistical analysis is done by moving a window of length $t$  along the sequences and counting how many times one finds $x$ markers inside this  window. Finally, $p(x;t)$ is obtained by dividing this number by the total number of windows of size $t$, namely by $N-t+1$, where $N$ is the total length of the sequence.
Since we typically deal with large values  of both  $x$ and $t$, we legitimately adopt a continuous approximation. Moreover, we make the ergodic and stationary assumption, thereby  expecting the emergence at long times of the scaling condition expressed by

\begin{equation}
\label{scaling}
p(x;t)=\frac{1}{t^{\delta}} F \left( \frac{x-wt}{t^{\delta}} \right),
\end{equation}
where $w$ is the overall probability of observing a marker, $\delta$ is  the {\em scaling index} and $F$ is a function, sometimes called ``master curve''. If $F$ is the Gauss function, $\delta$ is the known Hurst index, and if the further condition $\delta=0.5$ is obeyed, then the process is Poisson, the variable $x$ undergoes ordinary Brownian diffusion. According to Ref. \cite{katja} the occurrence of markers in time is regulated by random fluctuations, and the diffusion process occurs with no memory.

It straightforward to show that the Shannon Information

\begin{equation}
\label{shannon}
S(t)=\int_{-\infty}^{\infty}dx p(x;t) \ln p(x;t)
\end{equation}
with condition (\ref{scaling}) leads to

\begin{equation}
S(t)=k + \delta \ln t,
\end{equation}
where $k$ is a constant. The evaluation of the slope
according to which $S$ increases with $\ln t$ provides
therefore a measure for the anomalous scaling index $\delta$.

Let us briefly mention what we know about applying DE to time series with
non-Poisson statistics. We derive from the time series under study an
auxiliary sequence, by setting $\xi_i=1$ (this means that we find the marker
at the $i$th position), or $\xi_i=0$ (the $i$-th symbol, or value, is not a
marker). We then assume ``informativity'' for the markers. As explained in the
first part of Section 1, this means that we assume the markers to be crucial
events, and, that, consequently, the distance between a given ``1''' and the
next does not depend on the distance between this given Ò1Ó and the preceding
one. As mentioned earlier, if the lengths $t$
of the laminar regions are distributed as\begin{equation}
\label{model}
\psi(t) = (\mu - 1) \frac{T^{\mu-1}}{(t+T)^{\mu}}
\end{equation}
($\psi(t) \sim t^{-\mu}$ asymptotically is a sufficient condition),
then the theory based on CTRW and GCLT yields for $p(x;t)$ a truncated L\'evy
probability distribution function (PDF) \cite{giacomo}. 
DE detects the approximate
scaling $\delta$ of the central part, namely

\begin{equation}
\label{deltamu}
\delta=\frac{1}{\mu-1}  \:\: \mbox{if} \:
2<\mu<3,\:  \delta=0.5 \:\:  \mbox{if} \: \mu>3.
\end{equation}
We note that $p(x;t)$ is abruptly truncated by a ballistic peak at $x=0$
representing the probability of finding no marker up to time $t$. However, the
diffusion entropy increase is essentially determined by the occurrence of
crucial events, thereby leading to the scaling of Eq. (\ref{deltamu}).

It is worth explaining why the condition $2<\mu<3$ means long-range memory. The first, and probably easier way, is based on observing that
truncated L\'evy probability distribution $p(x;t)$ yields: 
\begin{equation}
\langle x^2 (t) \rangle - \langle x (t) \rangle^2 \propto t^{4 -\mu}
\end{equation}
thereby leading, as shown in Ref. \cite{elena} ,  to the 
correlation function

\begin{equation}\label{otto}
\frac{\langle (\xi (t_1)- \langle \xi \rangle) 
(\xi (t_2)- \langle \xi \rangle) \rangle}{\langle \xi^2 \rangle}
\propto \frac{1}{|t_2 -t_1|^{\mu -2}}.
\end{equation}
Note that the memory stemming from the non-Poisson nature of markers,
corresponds in this case to a non-integrable correlation function,
which is therefore a clear signal of infinite memory. In the
Poissonian case this correlation function would be exponential,
thereby indicating a much shorter memory.

There exists a second, and more impressive way, of relating the
non-Poisson distribution of waiting times to memory, according to the first
part of Section 1. We are observing the fluctuations of a dichotomous variable
$\xi$, with $\xi = 1$ implying the occurrence of a crucial event, and $\xi =
0$ the non-occurrence of events. 
If the role of time is played
by the ordinal number of words, we shall show that a good choice of
\emph{semantic} markers makes Eq. (\ref{model}) become a good model for the
dynamics of concept of natural language. The occurrence of Poisson statistics
would mean, according to Ref. \cite{katja}, that the time evolution of markers
is driven by fluctuations that are identical to white noise. In conclusion,
Poisson statistics would be totally incompatible with the existence of any
form of memory. The exponential decay of the correlation function of
Eq. (\ref{otto}) would not be a short-time memory, but it would signal the
complete lack of memory. The long-range correlation corresponding to $\mu <
3$, on the contrary, is a clear manifestation of the infinitely extended
memory of a text, associated to the strikingly non-Poisson character of the
waiting time distribution of Eq.(\ref{model}).

Eq. (\ref{deltamu}) \cite {giacomo} rests on uncorrelated waiting times
between events. This means that if two markers are separated by intervals of
words of duration $\tau_k$ (the distance in words between the $k$-th and the
$k+1$-th occurrence of the marker) then $\langle \tau_i \tau_j \rangle \propto
\delta_{ij}$, where $\delta_{ij}$ is the Kroeneker delta. Under these
conditions each event carries the same amount of information. The statistical
independence of different laminar regions means that the information carried
by the events is maximal for a given waiting time distribution
$\psi(\tau)$. In a linguistic jargon, we can say that if in a corpus we find a
marker (e.g. a list of words) such that $\delta \approx 1/(\mu - 1)$ then this
marker is {\em informative} in that corpus. If the markers are not
informative, this important condition is violated. We shall see hereby that
certain markers, e.g. punctuation marks, are not {\em real events}, but are
rather modeled by a Copying Mistake Map (CMM) \cite{maria}. This mean that
discourse complex dynamics are such that, although the punctuation marks carry
long-range correlations, and consequently the anomalous scaling of $p(x;t)$ ,
the waiting times in the laminar regions determined by these markers are
correlate. Punctuation marks are not informative. Their complexity is just a
projection of a complexity carried by ``concept dynamics'', namely, by the
really crucial events.

\subsection{The CMM and non-informative markers}

The Copying Mistake Map (CMM) \cite{maria} is a model originally
introduced to study the anomalous statistics of nucleotides dispersion
in coding and non-coding DNA regions. The CMM is a sequence that can
be derived from the long-range sequence generated by the waiting tiime
distribution of Eq. (\ref{model}), and corresponding to the long-range
correlation function of Eq. (\ref{otto}) as follows. We define the
probabilities $p(1) = \epsilon$ and $p(2) = 1 -\epsilon$. For any site
of the sequence, with label $i$, we leave the value $\xi_{i}$
unchanged with probability $p(1)$ and, with probability $p(2)$ we
assign to it a totally random value (copying mistake).

The resulting waiting time distribution undergoes an exponential decay. This
is so because the occurrence of a given $1$ depends on two distinct
possibilities: it can be an ``original'' $1$, or an original $0$ turned into
$1$ by a copying mistake. Formally, we write
\begin{equation}
\psi_{exp}(t)=\psi_{rand}(t) \Psi_{corr}(t)+ \Psi_{rand}(t) \psi_{corr}(t),
\end{equation}
where $\Psi(t)\equiv \int_{t}^{\infty}dt' \psi(t')$ and
$\Psi_{rand}(t)\equiv \int_{t}^{\infty}dt' \psi_{rand}(t')$,
and

\begin{equation}
\psi_{rand}(t)=\ln \left( \frac{2}{1 - \epsilon} \right) \cdot \left(
\frac{2}{1-\epsilon} \right)^{-t}.
\end{equation}
We note that $\psi_{rand}(t)$ and, consequently, $\Psi_{rand}(t)$
undergo an exponential decay.  So it does, in the asymptotic limit,
the function $\psi_{exp}(t)$. What about the DE curve? The theory
predicts \cite{scafettalatora} a random ($\delta=0.5$) behavior for
short times, a knee, and a slow transient to the totally correlated
behavior. An example of CMM is given by punctuation marks in Natural
Language.  We choose punctuation marks as markers for an Italian
corpus of newspaper and magazines, of more than 300,000 words length,
called {\em Italian Treebank} (hereafter TB).  We assign the value $0$
to any word that is not a punctuation mark, and $1$ when it is, namely
to full stops, commas, etc. A sentence like ``Felix, the cat,
sleeps!'' is therefore transformed into ``$0 1 0 0 1 0 1$''.

Figs. II (a,b and c) show that the choice of these markers yield a time series
with the typical properties of a CMM sequence. This means that the waiting
times $\tau_{k}$ are correlated. According to section 1A this means that
puntuation marks are not crucial events. Notice however that the use of the DE
method detects an anomalous scaling parameter $\delta$. This means that there
are hidden crucial events and that they generate an infinitely extended
memory.


\begin{figure}
\begin{center}
\label{fig2}
\Large
\vspace{50pt}{\bf\hspace{-9cm}a)\vspace{-50pt}} \\
\includegraphics[width=15 cm,height=8cm ]{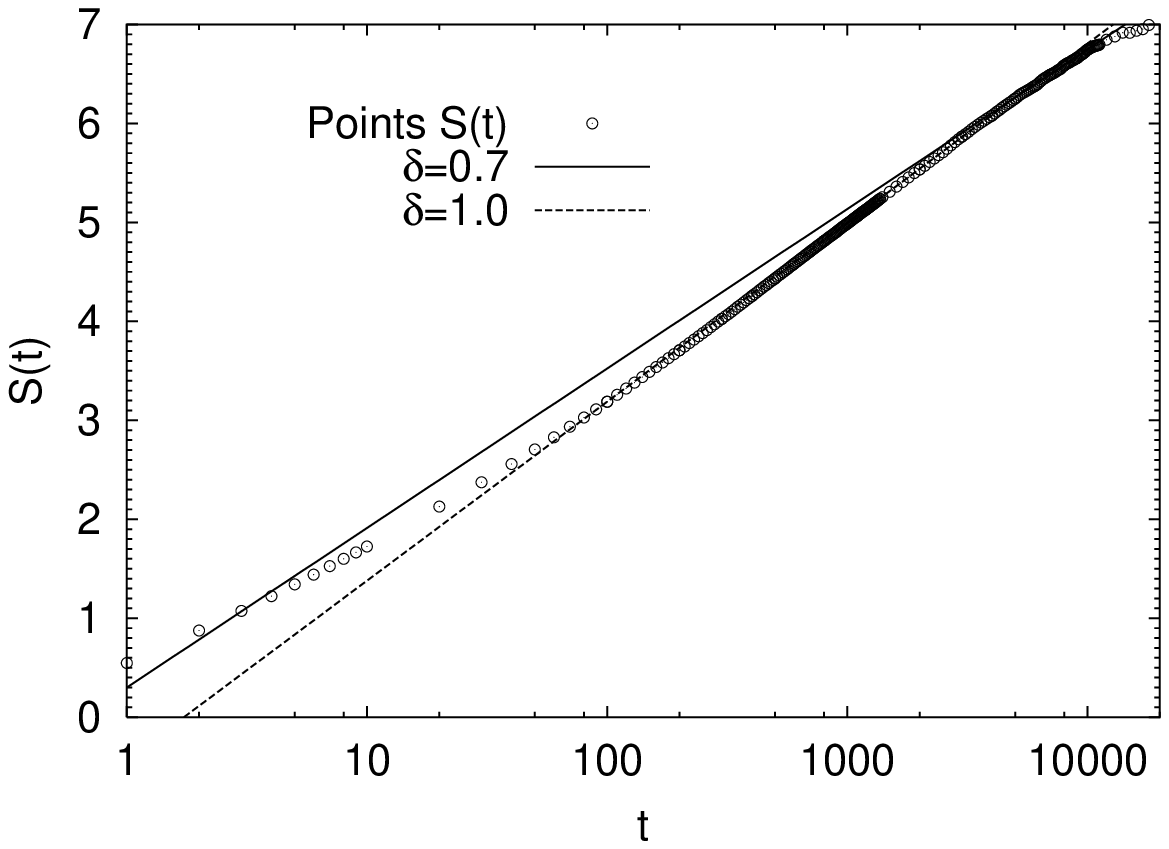}
\vspace{40pt}
{\bf\hspace{3cm}b)\hspace{8cm}c)\vspace{-40pt}}
\normalsize \\
\includegraphics[width=8 cm,height=6cm]{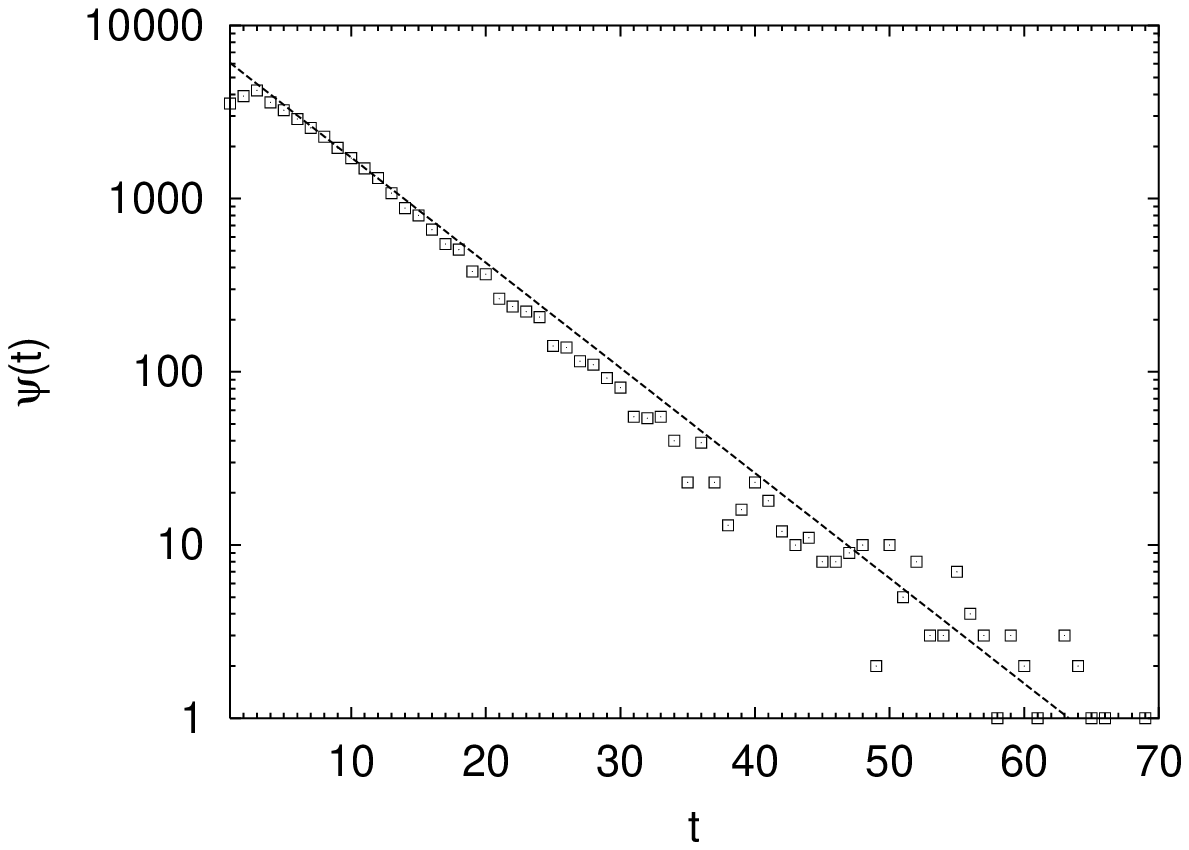}
\includegraphics[width=8 cm,height=6cm]{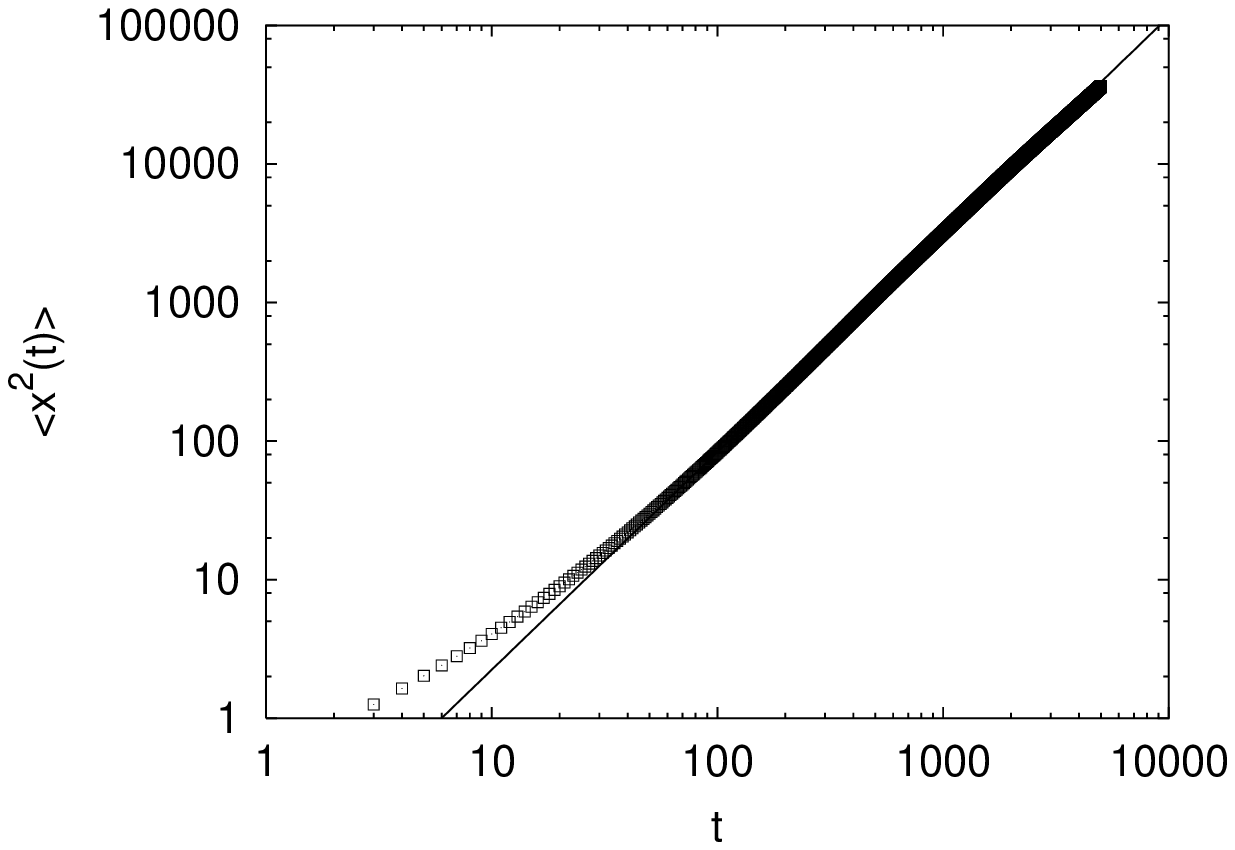}
\caption{a) Diffusion entropy for punctuation marks. The fit for the
asymptotic limit (solid line) yields a $\delta =0.7$. The dashed line
marks a transient regime with $\delta=1$ b) Non-normalized
distribution of waiting times for punctuation marks, namely counts of
waiting times of lenght t between marks in TB. The expression for the
dashed line fit is $7000 \cdot \exp (-t/7.15)$. c) Second moment
analysis for punctuation marks. The expression for the solid line fit
is $0.06\cdot t^{1.57}$. Notice that $1.57 \approx 3-1/0.7$, namely
the expression $H=3-1/\delta$ of Ref. \cite{scafettalatora} for L\'evy
processess stemming from CMM's is verified.}
\end{center}
\end{figure}



\subsection{Concepts as informative markers}

More experiments, not reported here, show that the CMM behavior is typical for
many characters, and are shared by all the letters of the alphabet, with a
$\delta \approx 0.6$. 
Obviously, If the marker is a given letter, the role
of time is played by the ordinal number of typographical characters in the
text. We note that in Italian the alphabetic characters mirror
phonetic. Thus, it is worth of linguistic interest to move from a phonetic to
a morphosyntactical level. To do so, a text has to be lemmatized and tagged
with respect to its part of speech. After this procedure we can identify as a
marker the occurrence of a certain part of speech e.g. article, adverb,
adjective, verb, noun, preposition, numeral, punctuation etc.). For instance,
the sentence ``Felix, the cat, sleeps!'' is now transformed into ``N P R N P V
P'', where N, P, R and V stand for nouns, punctuation, article and verb. If we
select a verb as a marker, we have ``$0 0 0 0 0 1 0$''. Fig 3 shows the result
of this experiment for verbs and for numerals. We notice that both cases lead
to the same asymptotic behavior of the DE, but to completely different
behavior of the waiting-times distribution $\psi(t)$ (where $t$ is the number
of words between markers). We notice that DE reveals a long-time memory in
both cases. As to $\psi(t)$, it shows an exponential truncation at long times,
in both cases. However, in the case of numerals we find an extended transient
with a slope $\mu_{numerals} < 2$, corresponding to the non-stationary, as
pointed out in the first part of the introduction. This is in fact due to the
uneven distribution of numerals in the corpus, since they are encountered more
often in the economic part of the Italian newspapers. As incomplete as this
analysis is, yet it reveals that numerals reveals carry more information than
phonetic markers or even than the verbs, used as markers. This is
linguistically interesting, since numerals denote a part of speech, but also a
``semantic class''.
\begin{figure}[hbt]
\begin{center}\vspace{40pt}
\Large
{\bf\hspace{3cm}a)\hspace{8cm}b)\vspace{-40pt}}
\normalsize \\
\includegraphics[width =8 cm,height=6cm]{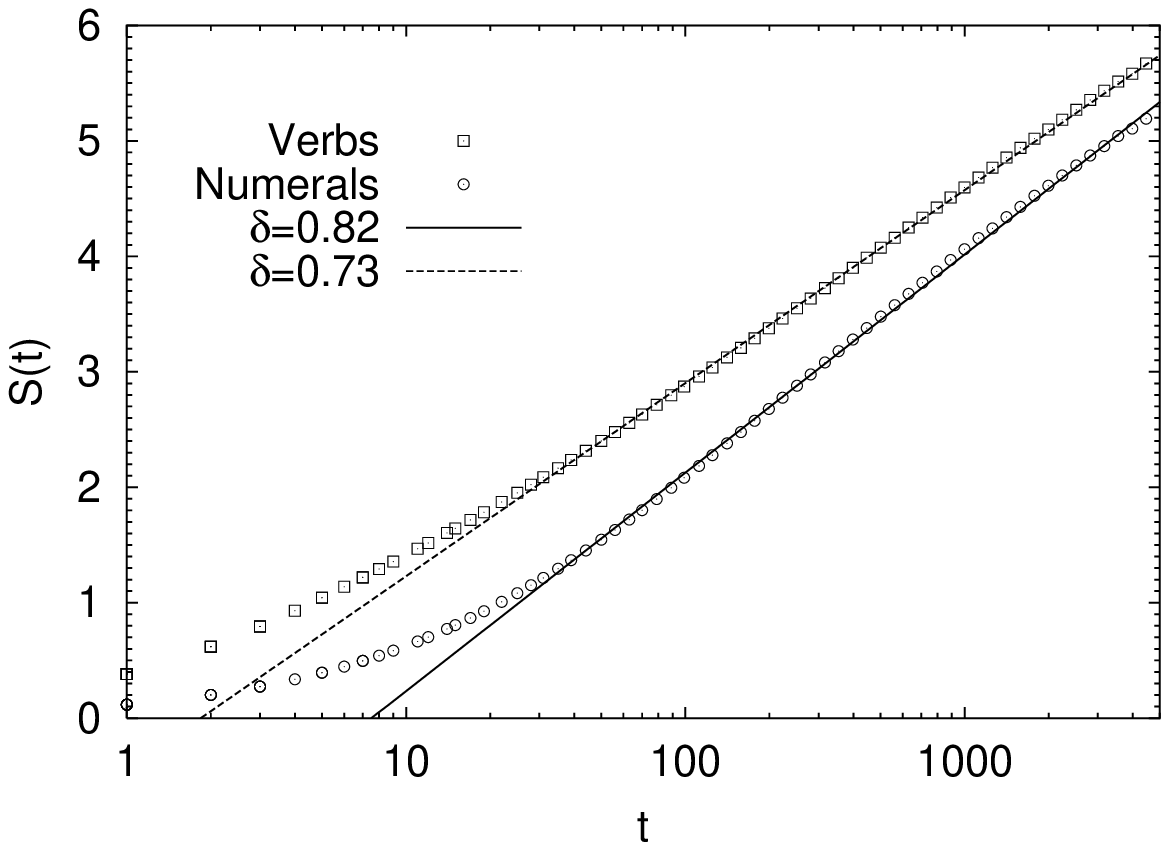}
\includegraphics[width= 8 cm,height=6cm]{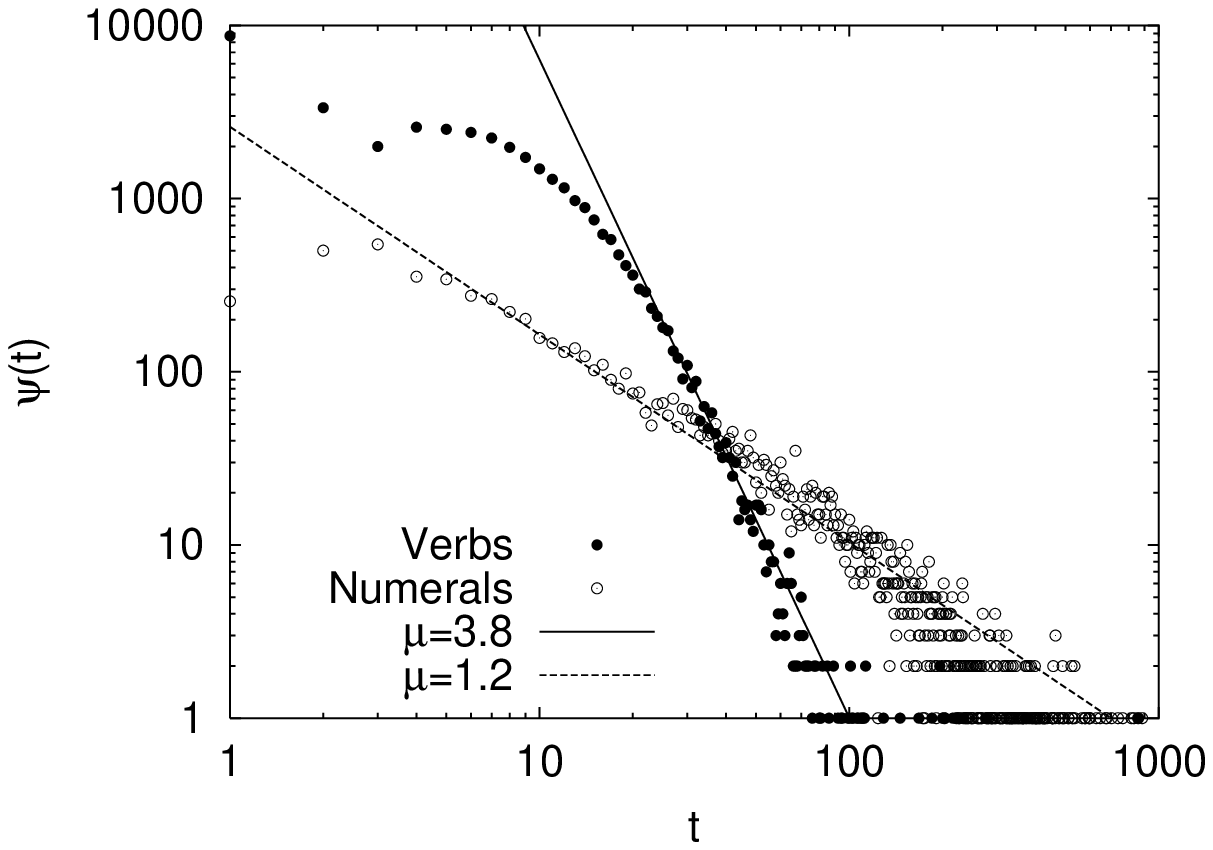}
\caption{a) DE for verbs (squares) and numerals (circles). The dashed
line is a fit for the verbs, with $\delta=0.73$, while the solid line
is a fit for the numerals, with $\delta=0.82$. b) $\psi(t)$ for verbs
(black circles) and numerals (white circles). The dashed line is a fit for the
numerals, with $\mu=1.2$, while the solid line is a fit for the
verbss, with $\mu=3.8$.}
\end{center}
\end{figure}


We are therefore led to suppose that informative markers are the ones
associated with a semantically coherent class of words. This is however a
problem, since every single concept is too rare in a balanced corpus (a long
text with a variety of genres). The next level of our exploratory search for
events is therefore to look for the occurrence of ``salient words'' in a
specialist text. Such a corpus has been made available as the Italian corpus
relative to the European project POESIA \cite{poesia}. POESIA is a European
Union funded project whose aim is to protect children from offensive web
contents, like, e.g. pornography in WWW URLs. Salient ``pornographic'' words
were automatically extracted by comparing their frequency in an offensive
corpus, to the frequency of their occurrence in the balanced TB corpus earlier
used. The definition adopted was

\begin{equation}
s(l)=\frac{f_{EC}(l)-f_{TB}(l)}{f_{EC}(l)+f_{TB}(l)},
\end{equation}
where $f_{EC}(l)$ is the frequency, in the erotic corpus, of the lemma $l$,
and $f_{TB}(l)$ is the same property in the reference Italian corpus (Italian
Treebank). Salient lemmata were automaticaaly chosen as the 5\% with the
highest value of $s$.  Notice that in this experiment all ``dirty'' words are
not taken into consideration, because they do not appear in the reference
corpus, and therefore $s=1$, but, in effect, cannot be properly defined,
especially for extremely rare words.
However an offensive
metaphoric use of terms is in fact detected, leading to a completely new way
to automatic text categorization and filter \cite{poesia}, using a method,
based on DE analysis, called CASSANDRA \cite{granada}.

Salient words were therefore used as markers for our analysis, as earlier
described. The results are shown in Fig. 4, clearly showing that {\em in a
specialized corpus, salient words of this genre, pass the test of
informativeness}. Salient words, and plausibly words in general, are therefore
distributed like markers generated by an intermittent dynamical model, with
$\mu \approx 2.1$ and, in agreement with (\ref{deltamu}), $\delta \approx
1/(\mu-1) = 0.91$. We see in the next section how this behavior is plausibly
connected with a topological complexity at the paradigmatic level, and in
Section III we derive the Zipf's law from the resulting model.

\begin{figure}[htb]
\begin{center}\vspace{40pt}
\Large
{\bf\hspace{3cm}a)\hspace{8cm}b)\vspace{-40pt}}
\normalsize \\
\includegraphics[width = 8 cm, height=6cm]{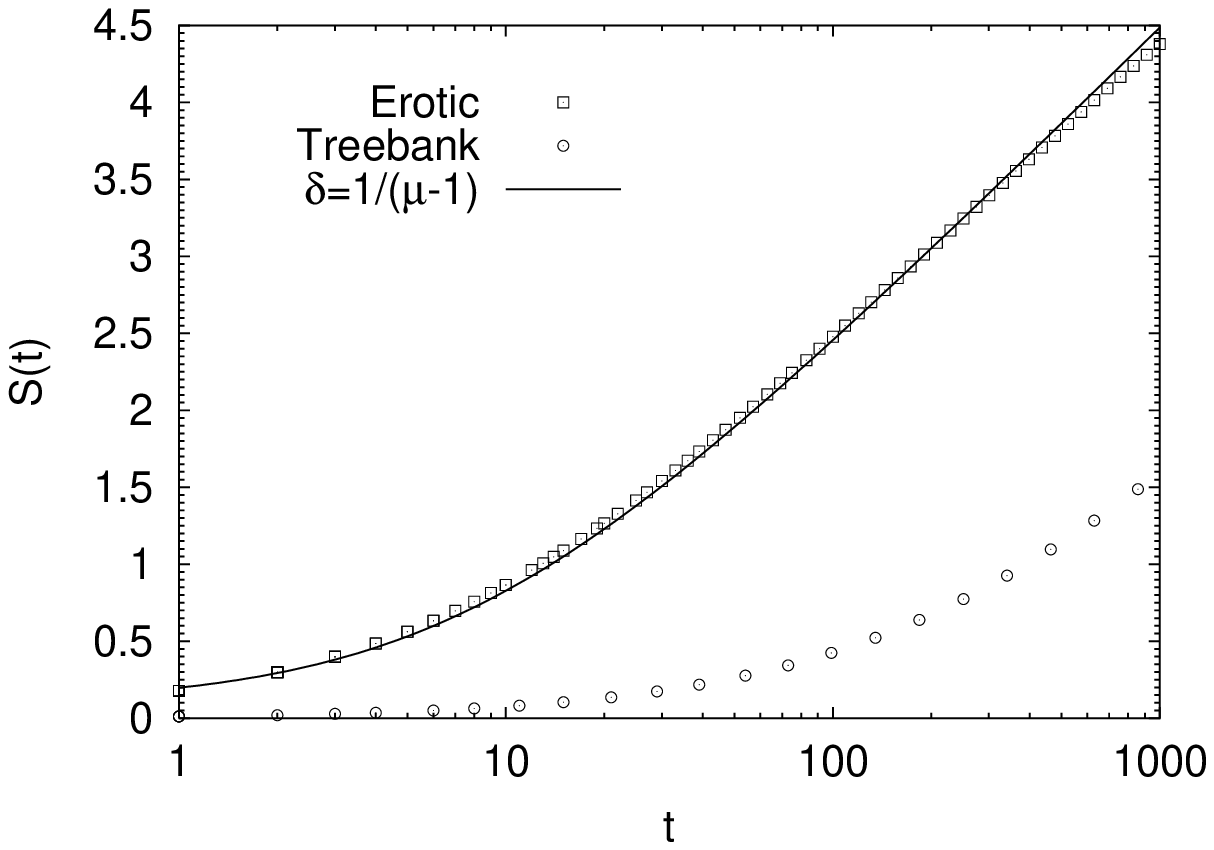}
\includegraphics[width = 8 cm,height=6cm ]{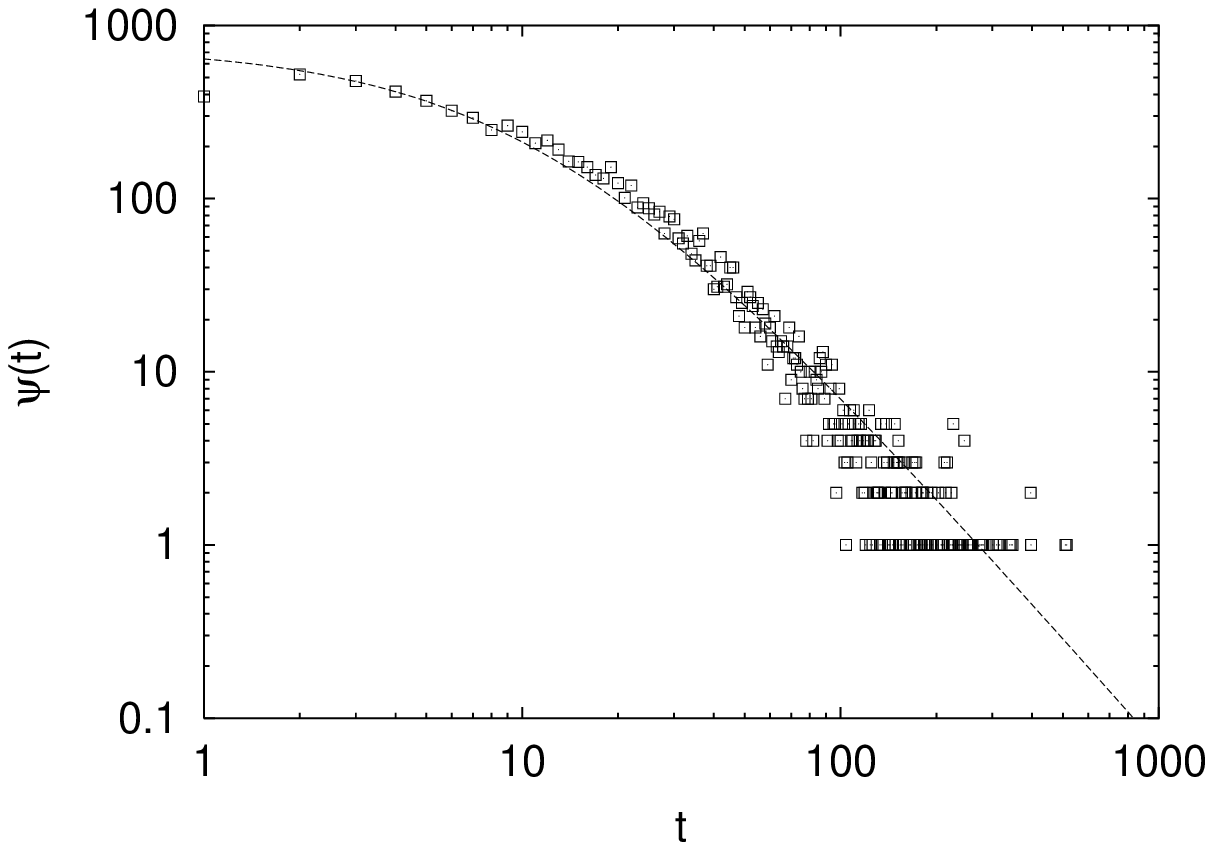}
\caption{a) DE for salient ``erotic'' words for a corpus of erotic
  stories and offensive web pages (squares), and for the italian
  reference corpus (circles). The solid line is a fit with expression
  $S(t)=k+\delta\ln(t+t_0)$, where the additional parameter $t_0$ is
  added to the original Eq. (3) to take transients into account and to
  improve the quality of the fit, yielding $\delta=0.91$ b)
  Non-normalized waiting time distribution for salient ``erotic''
  words for a corpus of erotic stories. The expression for dashed line
  fit is $14000\cdot(12.0 + t)^{-2.1}$, yielding $\mu=2.1$.}
\end{center}
\end{figure}



\section{Scale-free networks, intermittency and the Zipf's law}

In this section we outline a cognitive model that serves the purpose
of connecting structure and dynamics.  Allegrini et al. \cite{ilc}
identified semantic classes in the Italian corpus, by looking at
paradigmatic properties of interchangeability of classes of verbs with
classes of nouns. They defined ``super-classes'' of verbs and nouns as
``substitutability islands'', namely groups of nouns and verbs sharing
the properties that in the corpus you find each verb of the class co-
occurring, in a context, with each noun of the class \cite{ilc}. This
is precisely a direct application of the notion of ``paradigm''. Let
us call $p_v(c)$ and $p_n(c)$, respectively, the number 
of verbs or nouns belonging to a number $c$ of classes. They found that
\begin{eqnarray}
p_v(c) \propto \frac{1}{c^{1+\eta}}\nonumber \\
p_n(c) \propto \frac{1}{c^{1+\eta}},
\end{eqnarray}
where $\eta$ is a number whose absolute value is (much) smaller than $1$.

Working along the same lines, other authors \cite {smallworldenglish} found a
``small world'' topology \cite{smallworld}, by looking at the number of
synonyms in an English thesaurus, for each English lemma. We therefore assume
that this kind of structure is general for any language.  Let us therefore
imagine that the paradigmatic structure of concepts is a scale-free network
and consider a random walk in this ``cognitive space''.  Let us make the
following assumptions:

\begin{itemize}
\item The statistical weight of the $i$-th node is $\omega_i \sim c_i$
\item The dynamics of the system is ergodic so that the characteristic
recurrence time is $\tau_i \sim c_i^{-1}$
\item The above properties hold for all nodes, with the same functional form
for the recurrence time distribution, e.g. $\psi_i(t)=(1/\tau_i) 
\exp \left( -t/\tau_i \right) $
\end{itemize}
Now we imagine that selecting a {\it concept} means selecting a few
neighbouring nodes. This collection of nodes, due to the scale-free
hypothesis,
shares the same scaling properties
as the complete scale free network, namely $p(c) \sim c^{-\nu}$.
Therefore we have that
\begin{eqnarray}
\psi_{concept}(t)=\sum_{i} \omega_i \psi_i (t) \propto \sum_i c_i^2
e^{-c_i t} 
\approx \int dc  c^2 e^{-ct} \frac{1}{c^{\nu}} \sim
\frac{1}{t^{3-\nu}}.
\end{eqnarray}
Thus, we recover the 
intermittent model of Eq.(\ref{model}).

At this stage, deriving the Zipf's law $f \propto r^{-a}$, with $a$
close to unity, becomes a simple exercise Let us define frequency
probability $P(f)$

\begin{equation}\label{iacobian}
P(f)df = prob(r) dr \Longrightarrow P(f) \sim f^{-\frac{a+1}{a}}
\end{equation}
Next, let us notice that $P(f)$ must be a stable distribution. In
fact, the Zipf's law is valid for every corpus. In particular, if it
is valid for corpus $A$ and for corpus $B$, it is valid also for the
corpus $A+B$, where $+$ means the concatenation of corpora. If we keep
going on with this concatenating process, we shall have a corpus
\begin{center}
Total Corpus = Corpus A + Corpus B + ...   .
\end{center}
Thus,  the frequency of a word in the total corpus, $f_{tot}$,  is written
in terms of the single frequencies $f_1$, $f_2$, $\dots$, and of total
lengths $N_1$, $N_2$, $\dots$ of the single corpora, as follows
\begin{eqnarray}
f_{tot} = \frac{f_1+f_2+ \cdots }{N_1+N_2+\cdots} 
= \frac{1}{\sum_i N_i}\sum_i f_i, 
\end{eqnarray}
i.e., the Generalized Central Limit Theorem \cite{GCLT} applies.  This
means that the probability of frequency $P(f)$ is a L\'evy
$\alpha$-stable distribution. This probability of finding $f$
occurrences of a word in a corpus of a given length can be identified
with $p(x;t)$ of Section II, if we take into proper consideration the
length $t$ as a parameter. We have earlier noticed that $p(x;t)$ in
language is L\'evy process, with $\delta \sim 1$, and therefore with a
tail $P(f) \sim f^{-2}$. In other words through (\ref{iacobian}) we
recover (\ref{zlaw}) i.e. the Zipf's law.

\section{conclusions}

We have identified the cognitive process governing human language, and
proved it to be complex at both syntagmatic and paradigmatic
level. Although the study is conducted on Italian written corpora, we
are inclined to believe that the property found is language
independent, general and important, as emerging from a decade of
studies on the Zipf's law. Thus, we find that each concept corresponds
to a scaling close to $\delta = 1$, this being consistent with the
ZipÕs law. The complexity of language is located at the border between
the stationary ($\mu > 2$ and non-stationary ($\mu < 2$) condition
\cite{massi}. As pointed out in Section 1A, we adopt the perspective
of complexity as a condition of transition from dynamics to
thermodynamics. Within this perspective, we notice that, although the
condition $\mu > 1$ means aging \cite{jacopo} and a transition from
dynamic to thermodynamics in a virtually infinite time, a
thermodynamic condition exists, and diffusion process tends to
approach the scaling regime with $t \rightarrow \infty$. In the case
$\mu < 2$, no thermodynamic condition exists, and the stationary
condition cannot be realized, not even ideally. We see that specialist
texts are located within the ergodic regime ($\delta = 0.91$ in the
experiment illustrated in this paper). We are convinced that this is a
sign of the fact that language rests on the subtle balance between two
opposite needs, \emph{learnability}, which is the property consenting
a child to learn a language, and variability, namely the need of
exploring a virtually infinite cognitive space. Thus we are convinced
that the results of this paper might help to understand the language
complexity and evolution in children during the learning years, and in
psychopathological subject. Thus we propose its use for future
research work that might lead to the understanding of certain mental
diseases with no recourse to invasive diagnostic methods.

We think that the results of this paper might also help understanding
the complexity of really living systems. We have seen that a random
walk through the network, with the scale-free hypothesis generates
memory and long-range correlation. The speaker sets concepts in a
temporal order, but this corresponds to complex relations in a
non-linear space. We argue that a similar mechanism may also be chosen
by really living systems. Thus, the relational interconnectivity of
chemicals in the living cell may depend on the long-range memory of
the cell function, or influence it. From a Language Engineering point
of view, this study provides a theoretical background for a completely
new strategy of automatic text categorization. A prototype is being
implemented as a semantic filter \cite{poesia}. We think that the
proposed test for informativness for a set of markers will be
beneficial more in general for the analysis of time series.

Finally, we want to stress that the method successfully adopted in
this paper to establish the information content of markers, or labeled
values, of the time series, might have possible applications to the
war against terrorism. In fact, if the intelligence community affords
suggestions on the possible crucial words of a new wave of terrorist
attack, the adoption of this method allows an automatic way to
identify the messages with a terrorist content, more or a less in the
same way here illustrated to filter text with a pornographic content.

\emph{Acnowledgments} Financial support from ARO, through Grant
DAAD19-02-0037, is gratefully acknowledged.

\end{document}